\documentclass[submission,copyright,creativecommons]{eptcs}
\providecommand{\event}{ThEdu 2022} 
\newcommand{\PTB}{\textbf{Proof Tree Builder}}
\usepackage{graphicx}
\usepackage{iftex}
\usepackage{scalerel}

\ifpdf
  \usepackage{underscore}         
  \usepackage[T1]{fontenc}        
\else
  \usepackage{breakurl}           
\fi

\title{A Proof Tree Builder\\for Sequent Calculus and Hoare Logic}
\author{Joomy Korkut
\institute{Princeton University\\
Princeton, New Jersey, USA}
\email{joomy@cs.princeton.edu}
}

\newcommand{\titlerunning}{A Proof Tree Builder for Sequent Calculus and Hoare Logic}
\newcommand{\authorrunning}{J. Korkut}

\begin{document}
\maketitle

\newcommand{\plusbutton}{\texorpdfstring{\scalerel*{\protect\includegraphics{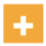}}{plus}}{plus}}
\newcommand{\minusbutton}{\texorpdfstring{\scalerel*{\protect\includegraphics{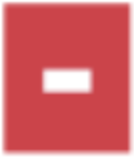}}{minus}}{minus}}
\newcommand{\detachbutton}{\texorpdfstring{\scalerel*{\protect\includegraphics{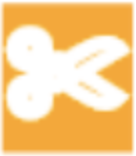}}{detach}}{detach}}
\newcommand{\hidebutton}{\texorpdfstring{\scalerel*{\protect\includegraphics{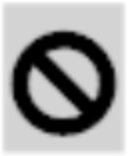}}{hide}}{hide}}

\begin{abstract}
  We have developed a web-based pedagogical proof assistant, the \PTB, that lets you apply rules upwards from the initial goal in sequent calculus and Hoare logic for a simple imperative language. We equipped our tool with a basic proof automation feature and Z3 integration. Students in the automated reasoning course at Princeton University used our tool and found it intuitive. The \PTB\ is available online at \url{https://proof-tree-builder.github.io}.
\end{abstract}

\section{Introduction}
\label{sect:introduction}

Producing proof trees is a tedious task in almost any form. When a student writes a proof tree on paper, they have no way to backtrack without using an eraser and slowing down the process. When a student types up a proof tree in \LaTeX, they must satisfy the compiler, at which point using the right inference rules becomes a secondary concern. Ideally, a program should make sure that the right rules are used and render the proof trees in proper \LaTeX, while the student focuses on the substance of the proof trees.

As a solution to this problem, we have designed a browser-based graphical proof assistant, the \PTB, which allows the user to construct proofs by specifying a proof goal, choosing the proof rule that should be applied next, and manually providing the necessary information for the rule to be applied (e.g.\ substitution terms, pre- and post- conditions). If a rule cannot be applied, the tool shows a warning message. The user is thus able to achieve a complete proof for the given proof goal by continuously applying proof rules. Experienced users can use the \textit{automation mode}, which hides irrelevant rules and has a basic automated propositional prover, while beginners can use the \textit{learning mode}, in which the user has to be more deliberate in the rules they pick.

The \PTB\ supports sequent calculus proofs~\cite{Szabo69} for first-order logic and Hoare logic proofs~\cite{hoare1969axiomatic} for a simple imperative language with sequencing, conditional, loop, and assignment statements. Students in the graduate-level automated reasoning course at Princeton University used our tool and found it intuitive, and our tool did not decrease their comprehension.


\section{A walk-through}
\label{sect:expf}

The \PTB\ interface has a top menu with buttons, and a left bar that has the list of active proofs, each of which has buttons to output the tree as \LaTeX, delete the tree, or save the tree as a file. The remaining white area is a workspace where proofs appear. Each proof tree can be dragged and dropped within the workspace. The user can also zoom in and out to focus on certain parts of bigger proof trees; the user experience is similar to that of a vector editing software.

\begin{figure}[!h]
  \centering
    \frame{\includegraphics[width=\textwidth]{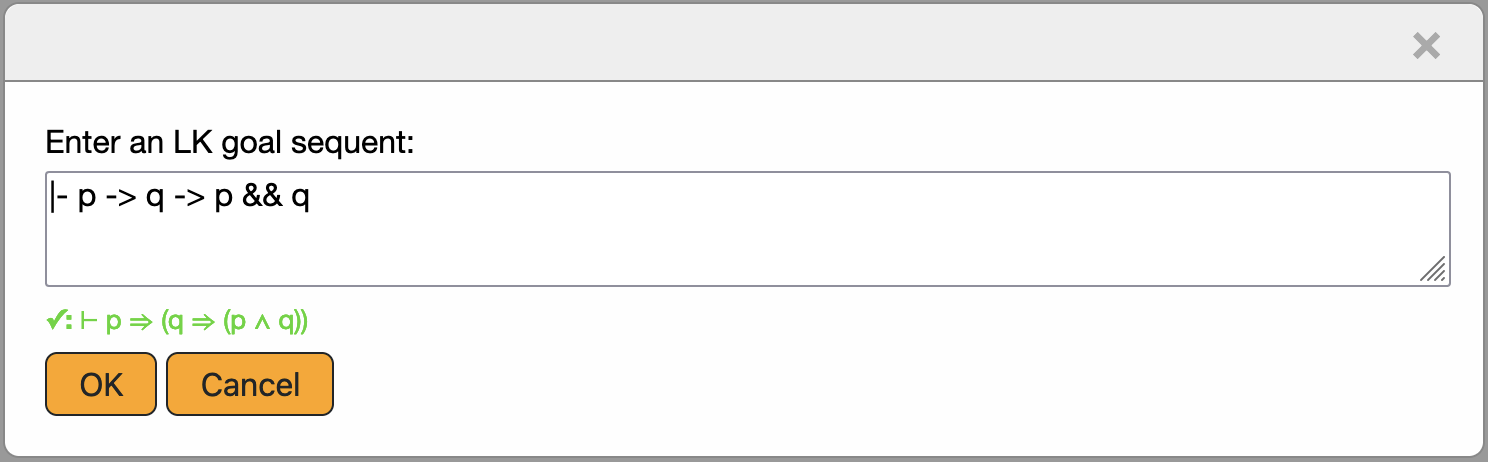}}
    \caption{Entering a goal in the \PTB.}
    \label{fig:parsing}
\end{figure}

Suppose the user wants to prove $\vdash p \Rightarrow q \Rightarrow (p \land q)$ in the learning mode. They would start by clicking the \textsf{Add LK Goal} button on the top left and then entering their goal, as seen in \autoref{fig:parsing}. The parser supports both Unicode characters such as $\land$ and ASCII representations of them such as \texttt{/\char`\\} and \texttt{\&\&}. As the user types, they can see how it is parsed below the text box. This feature helps the user double check the operator precedence and binder scopes. If there is a parse error, the user sees it below the text box in red.

\begin{figure}[!h]
  \centering
    \frame{\includegraphics[width=\textwidth]{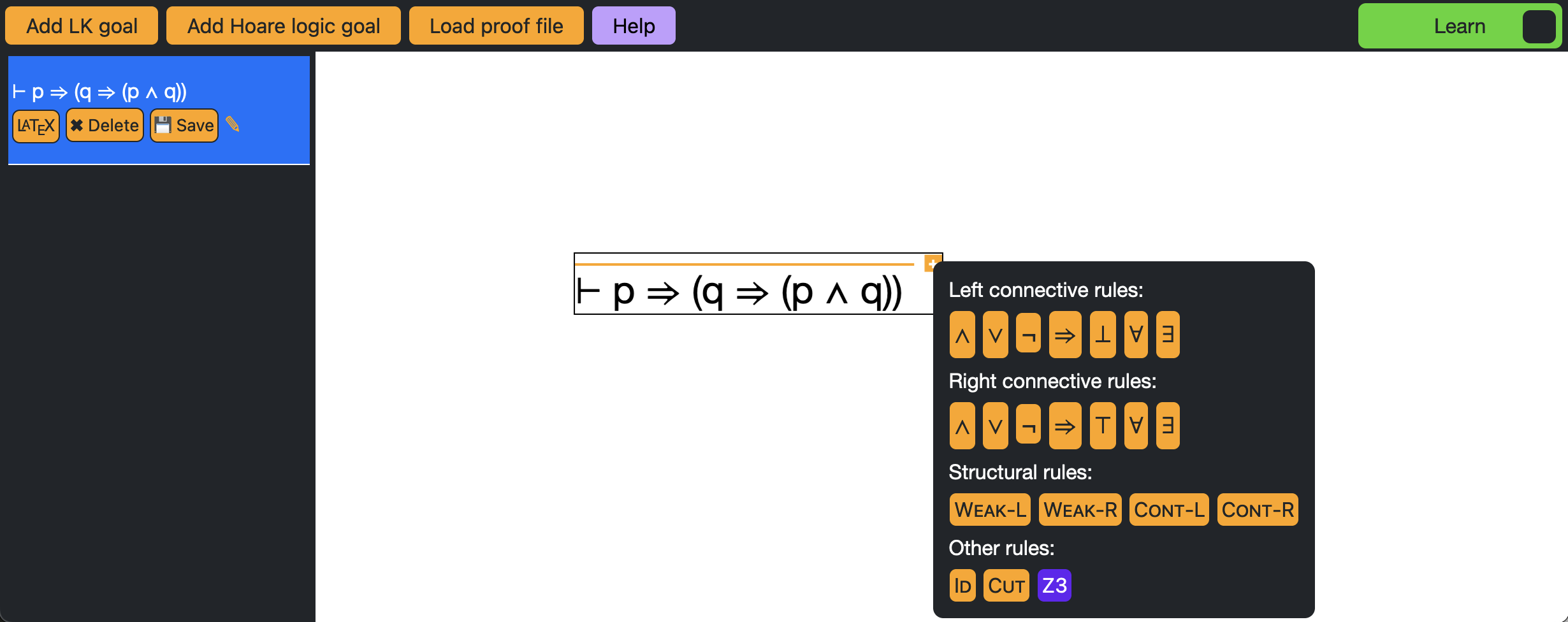}}
    \caption{Proof rule options popup menu after clicking the \protect\plusbutton{}\ button.}
    \label{fig:plusbutton}
\end{figure}

When the user adds a goal, they will see an incomplete proof tree in the workspace, as denoted by the orange line over the goal and the \plusbutton\ button on the right side of the orange line. When the user clicks on the \plusbutton\ button, a popup menu with all the proof rule options appear, as shown in \autoref{fig:plusbutton}.

\begin{figure}[!h]
  \centering
  \frame{\includegraphics[width=\textwidth]{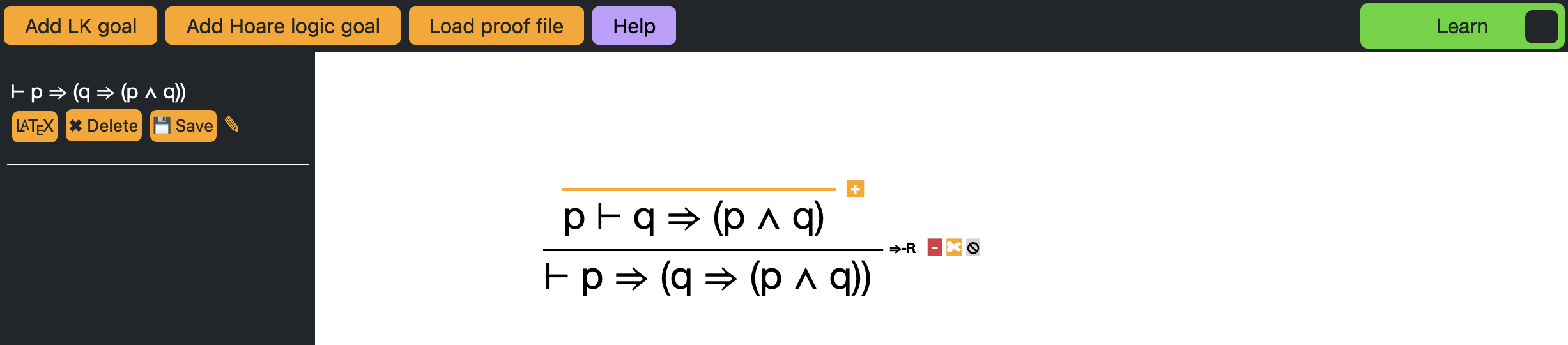}}
  \caption{Incomplete proof tree after one rule application.}
  \label{fig:addedonerule}
\end{figure}

\begin{figure}[!h]
  \centering
  \frame{\includegraphics[width=\textwidth]{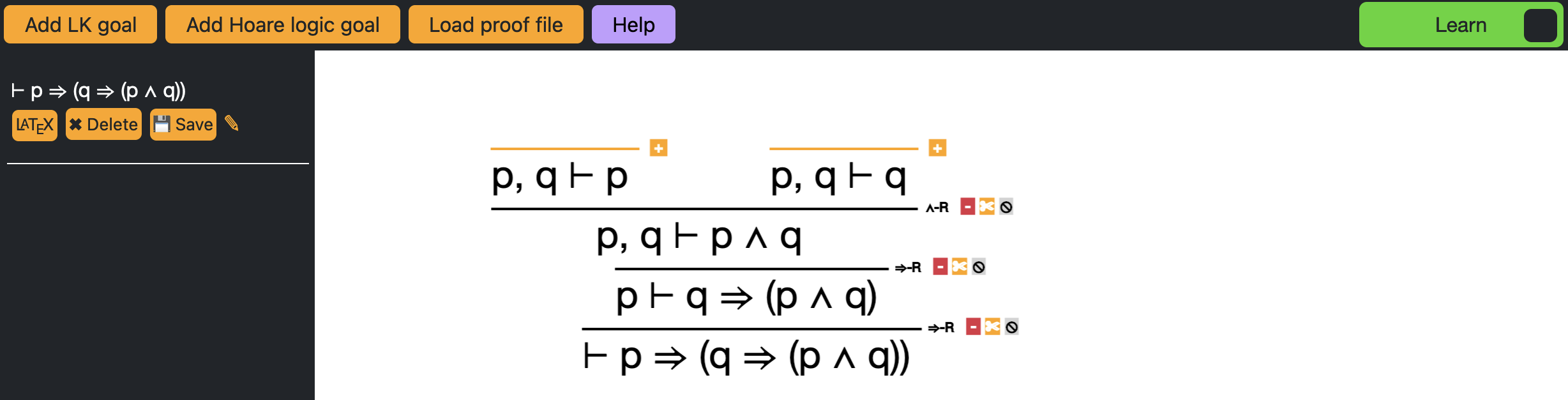}}
  \caption{Two incomplete premise proof trees.}
  \label{fig:twoincomplete}
\end{figure}

For the goal in the figure, the user needs to apply the $\Rightarrow_R$ rule, so they can click on the $\Rightarrow$ button under \textsf{Right connective rules} and reach the screen in \autoref{fig:addedonerule}.
The proof then proceeds with one more application of the $\Rightarrow_R$ rule and an application of the $\land_R$ rule. After the latter step, both premises of the rule will turn into incomplete proof trees, as seen in \autoref{fig:twoincomplete}. These incomplete proof trees should be completed separately by the user.
This proof can then be finished by applying the \textsc{Id} rule in both incomplete subtrees.

After a proof rule has been applied, the user sees three buttons, \minusbutton{}, \detachbutton{}, and \hidebutton{}, next to the rule label. These buttons can be used to undo proof rule applications. The first button, \minusbutton, unapplies the proof rule, and all the rules above. The second button, \detachbutton, divides a proof tree into two by detaching the subtree above into a separate proof tree, which makes the original proof tree incomplete. A detached proof tree can be dragged and dropped onto the incomplete end of another proof tree to reattach. The third button, \hidebutton, hides the premises of the proof rule and can be used to fold/unfold parts of a proof while working on large proofs.

\begin{figure}[!h]
  \centering
  \frame{\includegraphics[width=\textwidth]{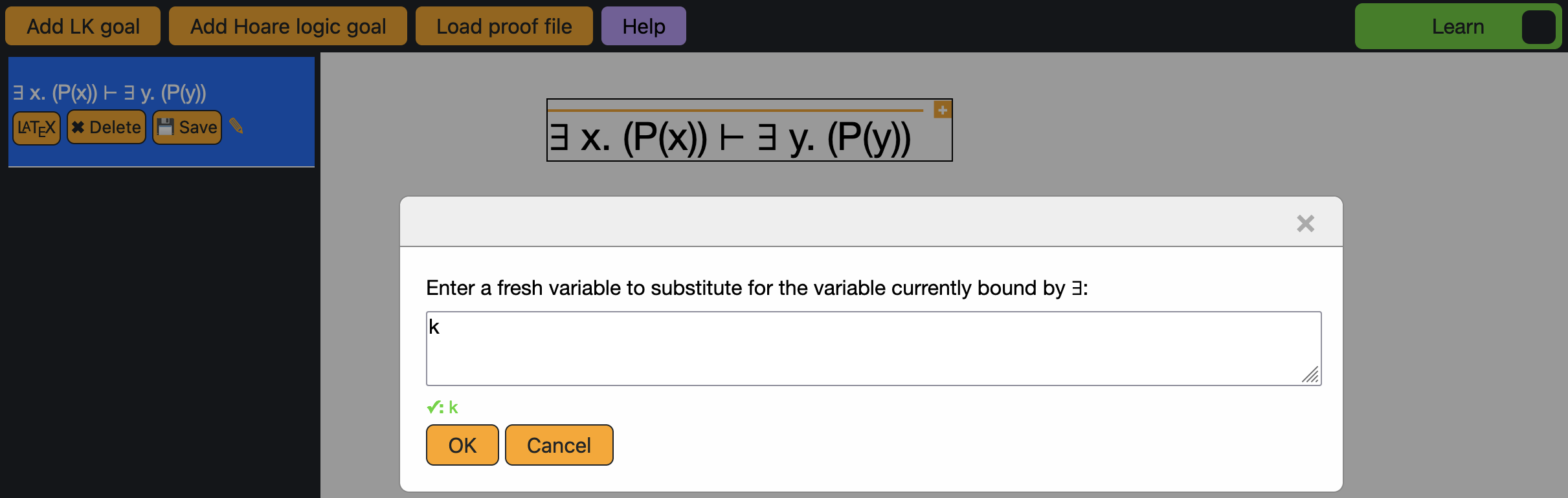}}
  \caption{A prompt asking the user for a fresh variable when applying the $\exists_L$ rule in a sequent.}
  \label{fig:firstorderk}
\end{figure}

For first-order logic sequents that contain quantifiers, clicking on the rules for the quantifiers will show the user a prompt asking for a fresh variable or a term, depending on the rule, as seen in \autoref{fig:firstorderk}.

\begin{figure}[!h]
  \centering
  \frame{\includegraphics[width=\textwidth]{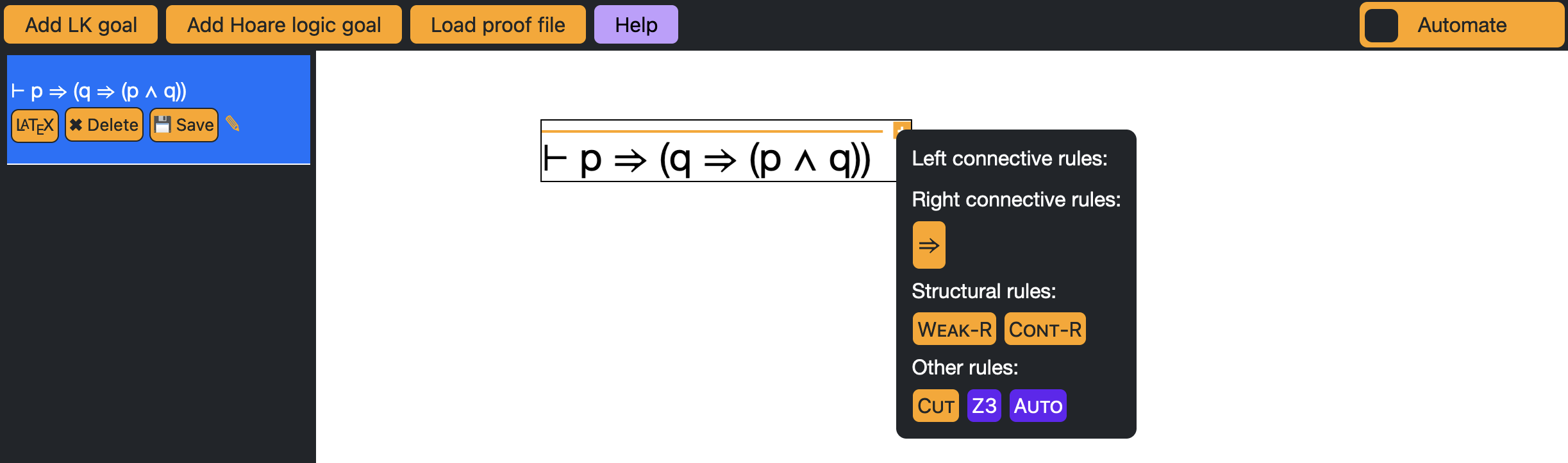}}
  \caption{The automation mode hides the buttons for the sequent calculus proof rules that are impossible to apply.}
  \label{fig:autobutton}
\end{figure}

The demonstration we have given above shows the \emph{learning mode} of the \PTB. The user can switch to the \emph{automation mode} by clicking the switch at the top right corner. In the automation mode, clicking on the \plusbutton\ button to apply a new rule will only show the relevant buttons. For a sequent calculus proof, the relevant buttons are the rules for the outermost connectives for both sides of a sequent, the structural rules depending on the number of formulas in each side of the sequent, the cut rule, and the automation buttons, as seen in \autoref{fig:autobutton}. For a Hoare logic proof, the relevant buttons are the rule for the immediate command and the logical rules.

In the automation mode, there is a new button, the \textsc{Auto} button, that does not appear in the learning mode. This button performs a primitive proof search. For the goal in \autoref{fig:autobutton}, the \textsc{Auto} button can actually generate the entire proof tree. \autoref{sect:automation} discusses the extent of its capabilities.

\begin{figure}[!h]
  \centering
  \frame{\includegraphics[width=0.5\textwidth]{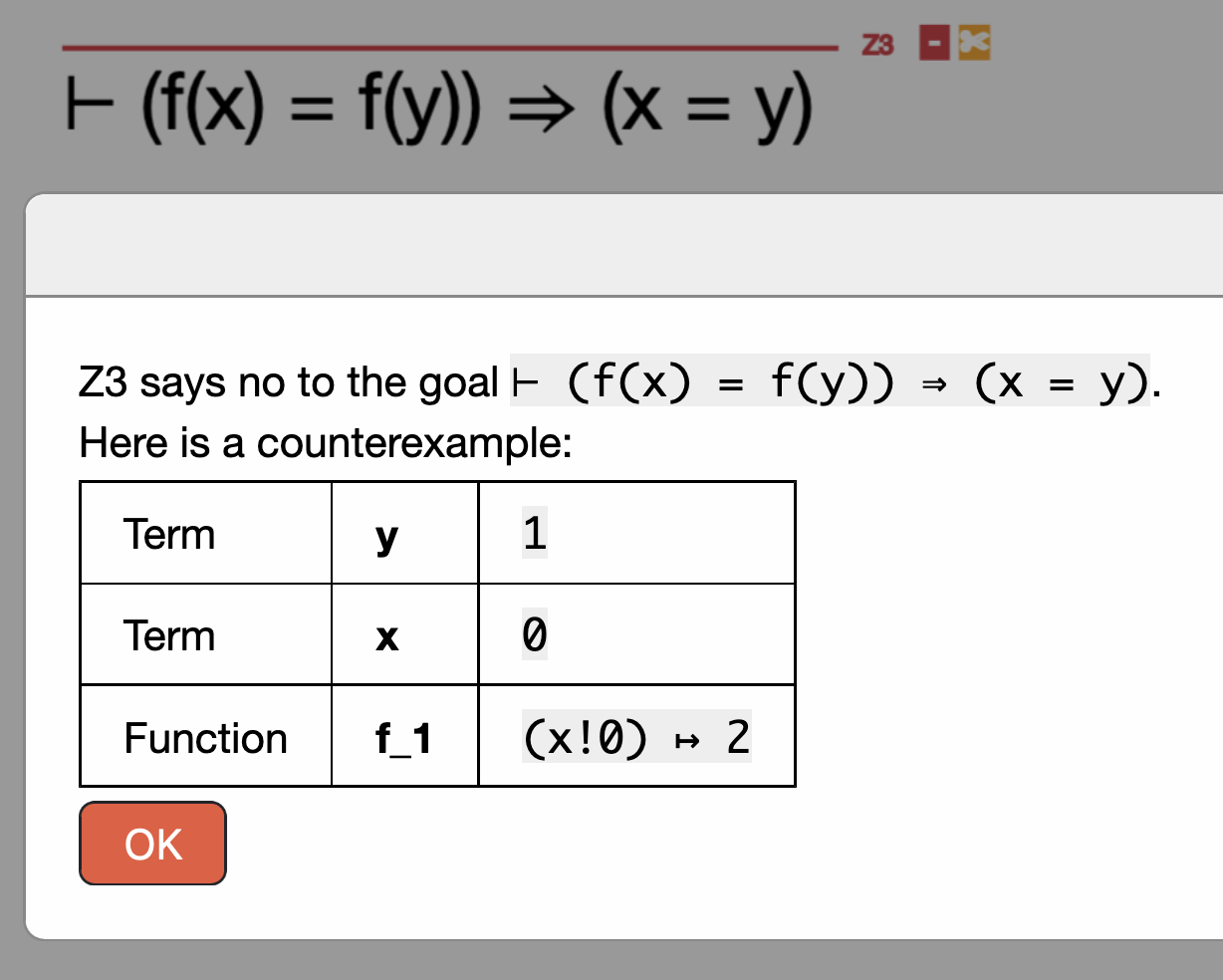}}
  \caption{Z3 generating a countermodel for the given goal.}
  \label{fig:z3counterexample}
\end{figure}

There is another button called \textsc{Z3}, which provides a different kind of automation. It runs the Z3 theorem prover~\cite{z3} on the goal, and if Z3 proves validity, our tool discharges the goal and draws a green line over the goal. If Z3 can find a countermodel, our tool shows it to the user, as seen in \autoref{fig:z3counterexample}.
\autoref{sect:z3} discusses this feature in more detail.

\begin{figure}[!h]
  \centering
  \frame{\includegraphics[width=\textwidth]{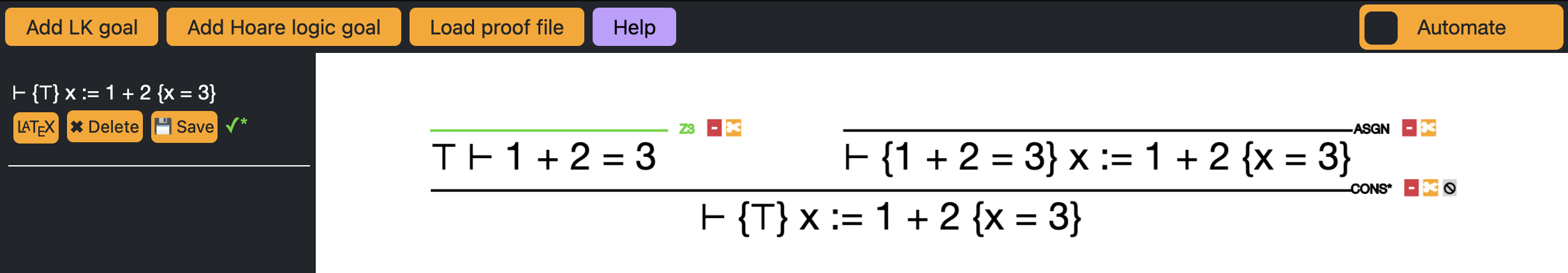}}
  \caption{Applying the Z3 pseudo-axiom in a simple Hoare logic example.}
  \label{fig:hoarez3}
\end{figure}

\autoref{fig:hoarez3} shows a simple Hoare logic proof with a consequence rule application. The first premise of the consequence rule is a sequent calculus goal inside a Hoare logic proof tree. However, our sequent calculus rules cannot interpret functions and relations on numbers, so the user has to use the Z3 pseudo-axiom to check the numeric equality goal.

\section{On design and implementation}
\label{sect:choices}

\subsection{Automation}
\label{sect:automation}

The automation button is not very clever; it merely applies available rules in a specific order. It gets stuck when a rule can be applied to multiple formulas in a sequent. For example, a propositional sequent such as $p \Rightarrow q, p \Rightarrow r, p \vdash q$ should be easily provable, but our prover needs to know which of the formulas the user would like to apply the $\Rightarrow_L$ rule to. This choice can lead to different proof trees, so our automation system does not make that choice, it just stops instead.

This limitation can be justified as a pedagogical choice: if the students can automate more of the proof, they may take that option more often and end up learning less.
There is value, however, in implementing more automation features as a way to provide hints to the user. We may choose to implement some of the existing work on such techniques~\cite{ehle, voronkov, wang} in the future.

\subsection{Z3 integration}
\label{sect:z3}


The homework assignments in our automated reasoning course~\cite{cos516} include writing proof trees for propositional and first-order sequents, and Hoare triples for programs that handle integer arithmetic. For sequents, proof trees are easily constructable, but Hoare triples involving integer arithmetic require proof rules for equality and arithmetic operations. Our theorem prover does have integers and arithmetic operators in terms, but it does not have proof rules for arithmetic function evaluation or the equality relation. We do not want to leave these goals unchecked. Therefore we added a new \emph{pseudo-axiom} to our tool for Z3. When applied, this rule encodes negation of the the sequent at hand in the SMT-LIB language~\cite{smtlib}, runs the Z3 theorem prover~\cite{z3} and looks for countermodels. If there are no countermodels, the goal is valid.

The Z3 pseudo-axiom is not a proof rule, it is merely a way for us to ``hand-wave" these proofs after checking if these sequents are valid modulo a theory. We make it clear to the user that this rule is not a formal proof rule; we are flouting the distinction between provability and validity on purpose. 

We made this pedagogical choice because adding all the rules for equality and arithmetic operations (and other operations if we extend with other data types) would have cluttered the proof assistant's interface and obscured the focus of these homework assignments. We would rather spend most of our time teaching the essentials of Hoare logic than get bogged down in details of equality and arithmetic operations.


The Z3 pseudo-axiom is also useful when the student is not sure if the goal they are trying to prove is valid in the first place. By applying the Z3 pseudo-axiom, they can find out if the goal is valid, and get a countermodel if it is not. If it is valid, they can always unapply the rule and continue to prove the sequent as usual.


\subsection{Implementation details}
\label{sect:implementation}

Our tool is written in JavaScript; it consists of static files and therefore runs solely on the client side. It uses an HTML5 canvas for the workspace, powered by the Fabric.js~\cite{fabricjs} library to handle graphics. The parser is generated from our grammar by PEG.js~\cite{pegjs}, a Yacc-style parser generator.
Our tool loads a version of Z3 compiled to WebAssembly via Emscripten~\cite{emscripten}. Z3 is then run by web workers, which makes the Z3 integration described earlier possible.

Formal guarantees about the correctness of our tool are not in the scope of this project. In our implementation, each rule is a class and the constructor of these classes check whether the rule application is correct. These checks can be inspected by reading the source code.

Saving a proof file creates JavaScript code that creates instances of the proof rule classes. When a proof file is loaded into \PTB, it evaluates this code and therefore reruns the correctness checks, which makes tampering with the proof files to get incorrect conclusions no more possible than getting an incorrect conclusion from within the tool.

\section{Related Work}
\label{sect:related}

Logitext~\cite{logitext} is one of the pioneers in web-based graphical proof assistants. It is a proof assistant for sequent calculus, where the user applies rules by clicking on a formula in the sequent. This interface can be confusing for new users, since their clicks change the proof state in cryptic ways. It can also be too easy once the user learns how to use the system, since they can keep clicking on random formulas until the proof is finished. Our system, in comparison, requires the user to explicitly pick a proof rule in individual rule steps.

The Sequent Calculus Trainer~\cite{ehle2015sequent, ehle2018sequent} is a tool developed at the University of Kassel, and it has similar features to the \PTB. We were not aware of their tool when we were developing ours, but we were pleasantly surprised to find similarities.
However, there are some significant differences. Their tool is written in Java and requires installation, while our tool is written in JavaScript and runs in the browser, with no need for installation.
Their tool only handles sequent calculus for first-order and propositional logic, while our tool handles Hoare logic as well. 
Both tools use Z3 to check for validity but our tool also presents countermodels for invalid goals to the user, as seen in \autoref{fig:z3counterexample}. In these respects, we believe our tool improves on their work.
The Sequent Calculus Trainer requires the user to pick a formula or subformula in the sequent and then to press a rule button. In contrast, our tool has adapted a different user interface to apply rules. Our tool has the \plusbutton\ button next to incomplete proof trees, inspired by the hole-driven development style of theorem proving of the ALF proof
editor~\cite{ALF}, Epigram~\cite{epigram} and Agda~\cite{agda}. Once the user presses \plusbutton\ and clicks on a proof, they do not have to pick which formula in the sequent the rule applies to, unless it is ambiguous, in which case they receive a prompt to pick which formula they mean. If a user of their tool picks a subformula inside a formula in the sequent and tries to apply a rule, they see a warning message that explains that rules can only be applied to top-level formulas. Their tool is pedagogically useful in the small, because it requires the user to have a better understanding of how the rules work, but ours is more pedagogically useful in the large, because it allows students to move fast enough to prove interesting theorems.

The Sequent Calculus Calculator~\cite{seqcalc} is another tool similar to ours. Their tool's user interface involves picking a rule to apply, filling the holes in the rule with formulas and terms that match how the user will apply the rule, and then dragging the rule on the part of the tree onto the part of the tree the rule should be applied. Their tool also handles first-order logic, but ours also handles Hoare logic for a simple imperative language. Our tool was first developed a year before this bachelor's thesis was written; we were not aware of their tool while developing ours. Their approach based on filling holes is a feature we would like to add to our tool as well, especially for steps like picking the middle formulas in a consequence rule application in Hoare logic, or for picking fresh variables or terms in quantifier rule applications in first-order logic.

There are pedagogical tools for building proof trees for other calculi as well. Panda~\cite{gasquet2011panda} is a proof assistant for natural deduction that allows building proof trees by backward or forward reasoning, with no automation, but with a similar graphical interface to the Sequent Calculus Trainer. ``Click and coLLecT''~\cite{callies2021click} is a tool for linear logic proofs, with almost the same interface as Logitext. Sequoia~\cite{sequoia} is a tool that allows various logic systems in sequent calculus style. Students using Sequoia pick an unproved sequent in the tree and pick a rule from the side bar to apply.

There is also a collection of tools for proofs in various calculi, in various proof styles other than trees.
NaDeA~\cite{nadea} is another natural deduction proof assistant; it lets the users write structured list style~\cite{kaye} proofs. Its syntax, semantics, and proof system are all formalized in the Isabelle proof assistant~\cite{nipkow2002isabelle}, so students can see ``behind the scenes", dig into the definitions all the way down, and verify the resulting proof in Isabelle. AXolotl~\cite{axolotl} is an educational tool that can handle Hilbert style, sequent style and natural deduction proofs. Their tool allows restricting the allowed rules for a given problem, hence helping the student focus on the rules that are relevant to the problem. Our tool's automation mode hides the buttons for the rules whose connectives do not match the formulas in the sequent, but it cannot hide rules based on their relevance to the entire proof. Carnap~\cite{carnap} is a framework that allows the users to define their own logic systems and write proofs in them, but not proof trees. Holbert~\cite{holbert} is a pedagogical proof assistant that allows users to define their own logic systems and also build proof trees using them. For a course that is not limited to sequent calculus and Hoare logic, such tools would be more useful. However, the generality of the such tools may cause the students to get less helpful error messages.

There are also projects that focus on a drag-and-drop model of building proofs~\cite{breitner2016visual, donato2022drag, lerner2015polymorphic}. Unlike our tool, these systems focus on gamification and appealing graphics.

\section{Conclusion}
\label{sect:conclusion}

The \PTB\ has been used in the automated reasoning course at Princeton University~\cite{cos516}. The students were given the options to either use the \PTB\ or write their proof trees by hand. Everyone used the \PTB\ and almost everyone got full points on the problems. There were no incorrect proofs, but some students incorrectly identified valid formulas as invalid and tried to produce a countermodel. The students did not ask any questions on the course discussion page about how to use the tool, which the professor thought was a data point in favor of our tool being easy to use. At first, the professor wondered whether the tool would ultimately decrease comprehension, since students can ``button-mash'' instead of understanding the calculus, but the results for relevant questions on the midterm exam were similar to the previous year's scores, indicating that student comprehension was intact. The professor reports that he plans to use the \PTB\ again next year. We have not done a further user study on the effectiveness of our tool.

The \PTB\ was born out of our frustrations writing proof trees in \LaTeX\ or on paper by hand. We resolved these frustrations by providing an intuitive interface to build proof trees for sequent calculus and Hoare logic and to learn these proof systems. We enhanced our system with automation features and we added Z3 support to confirm the validity of goals and to generate counterexamples for invalid goals.

\section{Acknowledgments}
\label{sect:acks}

I want to thank Anastasiya Kravchuk-Kirilyuk and John Li for their contributions to the development of the project, Anna Blech and Andrew W. Appel for their comments on the paper, and Zachary Kincaid for using our tool in his course.

\nocite{*}
\bibliographystyle{eptcs}
\bibliography{generic}
\end{document}